\documentstyle[11pt,aaspp4,tighten]{article}
\def\rosat{{\sl ROSAT~}}
\slugcomment{}
\begin{document}

\title{\bf Butcher \& Oemler Cluster A2111:\\
 A Head-on Merger at $z = 0.23$}
\author{Q. Daniel Wang and Melville Ulmer}
\affil{Dearborn Observatory, Northwestern University}
\affil{ 2131 Sheridan Road, Evanston,~IL 60208-2900}
\affil{E-mail: wqd@nwu.edu and ulmer@ossenu.astro.nwu.edu}
\affil{and}
\affil{Russell J. Lavery}
\affil{Department of Physics \& Astronomy, Iowa State University}
\affil{1 Osborn Drive, Ames, IA 50011-3160}
\affil{E-mail: lavery@iastate.edu}

\begin{abstract}
We present \rosat PSPC and HRI observations of A2111, the 
richest galaxy cluster photometrically surveyed 
by Butcher \& Oemler (1984). The cluster contains a distinct comet-shaped 
X-ray subcomponent, which appears hotter than the rest of the cluster. 
The galaxy and X-ray surface brightness distributions of the cluster show
a similarly elongated morphology. 
These results suggest that the cluster is undergoing a head-on 
subcluster merger. This merger may also be responsible for the high fraction of
gas-rich blue galaxies observed in the cluster. We have further detected 
a poor cluster along the merging axis and at a projected distance of 
$\sim 5$~Mpc from A2111.

\end{abstract}

\section {Introduction}

	It is becoming increasingly clear that clusters of galaxies 
are evolving even at relatively recent epochs. Analysis
of galaxy and X-ray surface brightness distributions (e.g., Geller \& Beers 
1982; Dressler \& Shectman 1988; Jones \& Forman 1992) indicate
that about 40\% of nearby clusters contain substructure. In particular, 
recent X-ray observations have revealed that 
A2256 and A754, two nearby rich clusters, are apparently undergoing major 
mergers with subclusters (Briel et al. 1991; Briel \& Henry 1994;  
Henry \& Briel 1995; Henriksen \&  Markevitch 1996). Therefore, at
least some of the clusters are dynamically young system. Furthermore, 
it has been known for a long time that
the average proportion of blue galaxies observed in rich clusters at  
$z \gtrsim 0.2$ is considerably higher than at lower redshifts
(Butcher \& Oemler 1984 --- BO hereafter; 
Lavery \& Henry 1986; Newberry et al. 1988; Dressler \& Gunn 
1992; Fabricant, Bautz, \& McClintock 1994). Yet unknown,
however, is whether or not there is a connection between the dynamic state 
and the blue galaxy population of a cluster 
(e.g., Couch et al. 1994; Zabludoff et al. 1996; Tomita et al 1996).

	The rich cluster A2111 provides an ideal site to explore 
this connection. The cluster 
is one of the richest clusters in the Universe. Among 33 rich clusters 
studied by BO (also see Butcher, Oemler, \& Wells 1983, BOW hereafter),
A2111 has the highest total number of cluster members brighter 
than $M_v = -20$ ($N_{30\%} = 155$). It also contains 
a large fraction of blue galaxies ($f_b = 0.16\pm0.03$; BO; Lavery \& Henry 
1986), higher than the average of clusters at similar redshifts ($z=0.2-0.3$), 
let alone that of nearby ones ($f_b \lesssim 0.05$). At its redshift of 0.228, 
A2111 permits a detailed study of both the X-ray-emitting intracluster medium 
(ICM) and the cluster galaxy population. 

	Based on two \rosat observations and
a deep R-band CCD image, we find that A2111
is most likely undergoing a head-on subcluster collision, and we argue 
that the collision may be responsible for the observed blue galaxy population.
In the following, we first
briefly describe the observations and the data reduction (\S 2),
and then present our results from a detailed analysis of the 
data, including a study of the galaxy distribution 
 (\S 3). We explore the implications of our results on both
the dynamic state and the galaxy population of the cluster (\S 4). 
Finally in \S 5, we summarize our results and conclusions. 
To be consistent with previous studies (e.g., BO),
we adopt $H_o = 50h_{50} {\rm~km~s^{-1}~Mpc^{-1}}$ and $q_o = 0.1$
throughout the paper. Thus, $1^\prime = 0.29 h_{50}^{-1}$~Mpc 
at the redshift of A2111. 

\section {Observations and Data Reduction}
 
	Our X-ray observations utilized the two X-ray detectors aboard 
\rosat (Tr\"umper 1992 and references therein): the Position Sensitive 
Proportional Counter (PSPC; Obs. No. rp800479)
and the High Resolution Imager (RHRI; rh800666).
The total exposures are 7,511~s (PSPC) and 62,287~s (RHRI). 
Briefly, the PSPC has a point 
response function (PSF) of $\sim 0\farcm5$ FWHM within the central 
$10^\prime$ radius and has about seven independent energy bands 
in the 0.1-2.4~keV range. The total field of view of the PSPC observation is 
2$^\circ$ diameter. But we use only the data within the central $16^\prime$ 
radius to avoid complications caused by both the detector's window supporting 
structure and the badly degraded PSF outside this radius (Hasinger et al. 
1992; Nichol et al. 1994). 
The RHRI observation has a higher spatial resolution 
of $\sim 5^{\prime\prime}$ on-axis.  However, the quantum efficiency
of the RHRI is about a factor of 3 lower than that of the PSPC, and 
the non-cosmic X-ray background is much higher (by a factor of 
$\sim 10$). Thus, the RHRI observation is useful primarily for resolving 
source confusion and bright emission peaks. 

	In our study of the diffuse cluster emission, we excise point-like 
X-ray sources, detected with signal-to-noise ratios greater than 3,
by excluding data within 90\% source flux-encircled radii. Compared to A2111,
all the X-ray sources in the vicinity (Figs. 1-2) are relatively faint.
In particular, within a 3\farcm5 radius of the cluster, we find no 
significant point-like emission. After rotating the RHRI image
by an angle of 0\fdg5 clockwise to correct for a systematic angular shift
(Kuerster 1993), we find good position coincidences (within a few
arcsec) between X-ray sources and optical objects from the APM catalog
(e.g., http://www.ast.cam.ac.uk/~lpinfo/apmcat/).
Also, the centroid of the only point-like X-ray source in the field of 
the R-band image (Fig. 3; see also Lavery \& Henry 1994)
coincides with a star-like optical object. The source positions 
of the two X-ray observations are also in good agreement. The source-excised
image is shown in Fig. 3; the residual source contamination in the image
is negligible in comparison to the statistical counting uncertainty. 

	We select data to maximize the signal-to-background ratio of A2111. 
The integrated light curve of the RHRI observation shows time intervals of 
enhanced contaminations from scattered solar X-rays. We exclude all 
those intervals of contaminations greater than the cosmic X-ray background, 
resulting in an adopted effective exposure of 40,080~s. We include 
data in the RHRI pulse-height channels 1-7 only. These selections together 
reduce the non-cosmic X-ray background level in the data by $\sim 70\%$. 
The non-cosmic X-ray background contamination in the PSPC observation 
is mostly negligible (less than a few percent), and 
we obtain an effective exposure of 7288~s. 
However, due to the emission from the Galactic hot interstellar medium,
the diffuse cosmic X-ray background is particularly bright in the soft band, 
defined here as the PSPC channel interval 20-41 
(sensitive to photons primarily in the 0.14-0.28~keV range). Our imaging 
analysis is therefore
chiefly in the hard band: 52-201 (0.5-2~keV). We further flat-field
the images, using exposure maps constructed in individual bands
(Snowden et al. 1994).
 
	Our study of the galaxy distribution in A2111 is based on the R-band 
image (Figs. 4-5), a median of three 600s exposures taken on the UH 2.2 
meter Mauna Kea telescope. Objects in the R-band image are morphologically 
classified
with the Picture Processing Package (Yee 1991). The classification
is complete down to 23.7 mag.  Between this limit and 17 mag, we find 
a total 777 galaxy-like objects (Fig. 6).

\section {Data Analyses and Results}

	Figs. 1-6 clearly show that A2111 has a complicated morphology. 
The cluster is strongly elongated at high X-ray intensity levels, 
and the degree of the elongation changes with intensity. The X-ray
emission peak is off-set from the centroid of large-scale low
surface brightness contours. The galaxy and X-ray distributions exhibit
a similar elongation. No correlation between 
individual galaxies and X-ray emission enhancements can be discernible by eye,
and an in-depth analysis is beyond the scope of this work. 
We now proceed to characterize both the spatial and spectral properties 
of the cluster and to examine its vicinity.

\subsection {1-D X-ray Surface Brightness Profiles}

	Although the morphology of A2111 clearly deviates from a circular 
symmetry, an 1-D characterization of the X-ray surface brightness 
distribution of the cluster can still be useful, especially
for comparisons with observations of more distant clusters. 
To calculate such a profile we first obtain a local ML (maximum likelihood)
centroid of the cluster X-ray emission peak within a 3\arcmin\ radius.
We choose this radius to avoid complications caused by both the
presence of discrete sources and nonuniform exposure of the observation at 
larger radii. The ML fit, similar to that described by Sarazin (1980),
uses the standard $\beta$ model of the form (Cavaliere \& Fusco-Femiano 1976):
\begin{equation} 
I = I_{o} \left(1+ {r^2 \over r_c^2}\right)^{1/2-3\beta},
\end{equation} 
where $r$ is the off-centroid radius. We conduct the fit in an iterative 
fashion. In each iteration, we fix the centroid position and use only PSPC 
counts within the chosen radius. Because the radius is considerably small
than the overall size of the cluster (e.g., Fig. 1), the fit does not 
provide tight constraints on the $\beta$ model parameters ($\beta,\ r_c,\ 
I_o$, plus a uniform background). But the fit does determine the local 
centroid relatively well. The best-fit position is at $15^h39^m40\fs9, 
+34^d 25^{\prime} 4\arcsec$ (R.A., Decl; J2000), and the 90\% uncertainty 
is about 6\arcsec. 

	Around the X-ray centroid, we calculate both the PSPC and RHRI 
surface brightness profiles, using source-removed count and exposure images.
Fig. 7 and Table 1 show the results of $\beta$ model fits to the profiles.
The model parameters obtained from the two profiles are consistent with each 
other. Evidently, the $\beta$ model provides a reasonably good description 
of the 1-D profiles. In particular, the profiles exhibit no 
enhanced central emission (above the $\beta$ model), therefore no sign 
of a cooling flow in A2111. The 1-D characterization, however, is not
sensitive to the 2-D substructure, which is apparent in the X-ray images.

\subsection {2-D Morphology}

	We  characterize the 2-D diffuse X-ray morphology of A2111
with a series of ellipses on various scales (Fig. 8).
Each ellipse is determined by four parameters: the center coordinates, 
ellipticity ($\epsilon$), and orientation of the major axis
($\theta$; north to east). We employ a simple moment method to calculate the 
parameters. This method was devised by Carter \& Metcalfe (1980) 
to study the galaxy distribution of clusters. 
McMillan, Kowalski, \& Ulmer (1989) and Buote \& Canizares (1994)  
applied the method to X-ray images of nearby clusters. We compute 
the required first and second order moments, 
using positions of individual counts without binning, thus utilizing
the full resolution of the data. Also because
the detected number of counts per resolution 
element is much smaller than one (e.g., Fig. 3), the computation based on 
individual counts are more efficient than on image pixels.

Our computation follows an iterative procedure (see also 
Carter \& Metcalfe 1980; Buote \& Canizares 1994).
Starting with a circular region, we calculate the moments. 
We then derive the center coordinates, ellipticity, and orientation 
from the moments. These parameters, together with the semi-major axis,
define an elliptical region. So we can return to calculate the moments
and repeat the process until the changes in the computed parameter values 
all become less than $10^{-3}$.

	In each iteration, we explicitly subtract the expected background 
contributions to the moments.  The computation of these background
contributions is analogous to the calculation of the 2-D moments
of inertia within an elliptical region of a uniform mass distribution.
We estimate the background in a source-subtracted region between 
10$^\prime$-16$^\prime$ off-cluster, where 
cluster contribution is negligible. Small uncertainties ($\lesssim 10\%$) 
in the background estimate may change the derived parameters by
a few percent, well within statistical errors.

	Proceeding from a large radius to small ones, we
obtained the parameters of the ellipses with their semi-major axes
from 1$^\prime$ to  8$^\prime$. The parameters 
on different scales are not totally independent, because 
the computation uses the {\sl aggregate} count distributions; on each scale, 
the parameters are average values 
within an ellipse, though weighted heavily by the outer parts of the region.

	We estimate the uncertainties in the parameters, using a
bootstrap algorithm. First, we generate 1000 bootstrap realizations by 
randomly re-sampling among count positions with replacement 
(e.g., Efron \& Tibshirani 1993). Second, we apply the above moment
method to each 
realization to obtain bootstrap replications of the ellipse parameters. 
Third, we sort the replications according to their values and
use the 5\% and 95\% percentiles of the values 
as estimates of the 90\% confidence interval of each parameter at a 
given semi-major axis. 

	Fig. 9 summarizes the results. The centroid coordinates of the cluster
clearly shift with semi-major axis, caused primarily by the off-center 
emission peak. The global centroid, moment-weighted within the largest 
ellipse, is $\sim 13^{\prime\prime}$ northwest 
of the northern most of the two major galaxies. This galaxy (BOW \# 6), 
at $\approx 15^h39^m40\fs4, +34^d 25^{\prime} 27\arcsec$, has long been 
identified in the literature as the optical center of the cluster
(Sandage, Kristian \& Westphal 1976). The ellipticity also changes 
significantly (by a factor of $\sim 2$).
The cluster is strongly elongated at a radius $\lesssim 
6^\prime$ and becomes more circular on larger scales. The variation in 
the position angle, though significant, is $\lesssim 20^\circ$. These 
variations are, however, only a conservative 
characterization of the true morphological distortion of the cluster,
because the parameters on different scales are correlated. 
 
	A similar analysis with the RHRI data at radii 
$\lesssim 4^\prime$, where the signal-to-background ratio is 
relatively high, gives results that are statistically 
consistent with those from the PSPC data.

	The galaxy distribution also shows an elongated morphology (Fig. 6).
The distribution, characterized with an ellipse of a  
semi-major axis $3^\prime$, has the centroid, ellipticity,
and orientation as 
$15^h39^m40\fs6 (-25\arcsec,+31\arcsec), +34\arcdeg25\arcmin27\arcsec
(-50\arcsec,+30\arcsec)$,
0.65(0.24-0.78), and 151(120-165) deg. The uncertainties are all at
the 90\% confidence level. These parameters are  
consistent with those from the X-ray data. Furthermore, the orientation 
between the two central major galaxies (BOW \#5 and \#6; Fig. 6) coincides 
well with the elongation of both the galaxy and X-ray distributions. The two
galaxies are 45$^{\prime\prime}$  apart ($\sim 220 h_{50}^{-1}$~kpc
in projection) and are nearly identical in color and magnitude.

\subsection{Morphological Decomposition}

	We approximate the X-ray emission of A2111 as a superposition
of a ``subcomponent'' on a large-scale ``main component''.
Assuming that the subcomponent is responsible for the off-center peak, we 
define the main component in the relatively feature-free northwestern half of 
the cluster between $-107^\circ$ and $73^\circ$ (north to east). 
We calculate the PSPC surface brightness profile 
of the main component as a function of 
$r^2=(-x {\rm sin}\theta+ y {\rm cos}\theta)^2+(x {\rm cos}\theta+ y 
{\rm sin}\theta)^2/(1-\epsilon)^2$,
expressed in the image coordinates ($x, y$) relative to the global
centroid of the cluster. The coordinates of the centroid 
($15^h39^m39\fs9, +34^d 25^{\prime} 39\arcsec$) as well as
$\epsilon$ (0.17) and $\theta$ ($167^\circ$) are from 
the ellipse fit with the semi-major axis equal to $8\arcmin$ (Fig. 9).
We fit the profile using the $\beta$ model as in Eq. (1), except that
$r_c$ is now the semi-major axis of the core. The best-fit is presented in 
Fig. 10, and the model parameters are $\beta = 0.65 (0.54-0.84), r_c = 
0.56(0.39 - 0.82)$~Mpc, and $I_{o} = 9.5(7.7-11) \times 10^{-3} 
{\rm~counts~s^{-1}~arcmin^{-2}}$. We then subtract this elliptical $\beta$ 
model, extended two-dimensionally to the whole 
region in Fig. 3, from the surface brightness distribution of A2111.

	The residual map in Fig. 11 demonstrates that the subcomponent is
apparently a coherent structure. There are three distinct features: 1) a 
dominant blob centered between the two major galaxies; 2) a long, possibly 
twisted tail to the northwest; 3) a lobe to the southeast. Integrated over 
the region enclosed by the contour of $3 \times 10^{-4} 
{\rm~counts~s^{-1}~arcmin^{-2}}$, the total count rate of the subcomponent
is $\sim 0.056 {\rm~counts~s^{-1}}$, corresponding to an absorption-corrected
luminosity of $\sim 2 \times 10^{44} h^{-2}_{50}{\rm~ergs~s^{-1}}$ 
in the 0.1-2~keV band (\S 3.4). About 60\% of this luminosity arises in the
 blob of $\sim 3^\prime$ diameter. The over-subtraction in regions near 
the blob (Fig. 11)
is, at least partly, due to the presence of the subcomponent tail. As 
the main component is a fit to the data that includes the 
tail, excluding the tail contribution from the fit would lead to an even 
flatter main component, and would, in turn, make the subcomponent even more 
prominent.

	The above choice of the global moment-weighted X-ray centroid as the 
center of the main component is logical, but not unique. Alternatively, we may 
choose the central galaxy, which is about 13\arcsec~ southeast. This choice
produces no qualitative change ($\lesssim 10\%$) in the 1-D profile of 
the main component, and slightly enhances the presence of the subcomponent 
tail. We conclude that the very presence and morphology of the 
subcomponent are due to both the ellipticity variation 
at different intensity levels and the off-center morphology of the cluster.

\subsection {X-ray Spectral Characterization}

	The limited spectral resolution and coverage of the PSPC data
allow only a crude X-ray spectral characterization of A2111. We collected an 
on-cluster spectrum in a circle of 5$^\prime$ radius around the cluster's 
local ML centroid (\S 3.1), and a background spectrum 
in a concentric annulus between 
10$^\prime$ and 16$^\prime$ radii. The cluster contribution to the background
in this annulus is negligible; different choices of the radii within the
range between  8\arcmin\ and 18\arcmin\ results in no discernible change in
the spectrum. The spectra are binned into the standard SASS channels
(Snowden et al. 1994).
We analyzed the spectra 
with the XSPEC package (e.g., http://heasarc.gsfc.nasa.gov/docs/xanadu/), 
assuming the Raymond \& Smith thermal plasma model
for the ICM emission. The spectra place only a 95\%-confidence 
lower limit on the metal abundance of the ICM as $\gtrsim 22\%$ solar. 
We thus fixed the abundance to be 30\% solar. Because the X-ray emission
is dominated by thermal bremsstrahlung, the uncertainty in
the abundance (within a factor of $\sim 2$; Mushotzky et al. 1996 
and references therein) has little effect on the model fits (a few percent
changes in the best-fit parameters). We further fix the abundance 
in X-ray-absorbing gas to be solar.

	The best fit with $\chi^2/d.o.f = 28.6/28$ is satisfactory (Fig. 12).
The measured temperature and absorbing-gas column density are
3.1(2.1-5.3)~keV and $2.2(1.8-2.6) \times 10^{20}{\rm~cm^{-2}}$;
the parameter intervals are all at the 90\% confidence. The column density
agrees well with the Galactic atomic hydrogen column 
density of $1.9 \times 10^{20}{\rm~cm^{-2}}$ from a 21~cm survey
(Stark et al. 1992). With the best-fit parameters,
we can convert the net cluster count rate $0.215\pm0.007 {\rm~counts~s^{-1}}$ 
within the 5\arcmin\ radius into an unabsorbed luminosity of $8.5 
\times 10^{44} h^{-2}_{50}{\rm~ergs~s^{-1}}$
in the cluster's rest-frame energy range 0.1-2~keV, or into a 
bolometric luminosity of $1.3 \times 10^{45} h^{-2}_{50}{\rm~ergs~s^{-1}}$.
The uncertainty is about 20\% in the 0.1-2~keV luminosity and is
up to about 40\% in the bolometric luminosity, within the quoted 90\% 
limits of the spectral parameters.
Similarly, we obtain an unabsorbed 0.1-2~keV flux/count-rate 
conversion of 5 $\times$ and 1.5  $\times 10^{-11} 
{\rm~ergs~s^{-1}~cm^{-2}/(counts~s^{-1})}$ for the RHRI and the
PSPC hard band, respectively.  The flux is again calculated in
the cluster's rest-frame 0.1-2~keV. 

To find out whether or not the subcomponent is spectrally different 
from the rest of the cluster, we further split the above on-cluster
spectrum into two: one from the dominant subcomponent region of 1\farcm2 
radius around the local ML centroid (\S 3.1), and the main component 
from the rest of the cluster. The radius is chosen so that the two spectra
have approximately the same signal-to-noise ratios. The contamination
between the two components tends to reduce the actual spectral difference.

	We fit the spectra jointly. 
Assuming that the absorptions in the two regions are the same, we obtain the 
sub- and main-component temperatures (keV) as 4.4 (2.2-15) and
2.7 (1.8-4.5). This temperature difference becomes a little more 
significant (at $\sim 95\%$ confidence), if we fix the absorption to 
the 21~cm measured Galactic value. Therefore, the average temperature
of the ICM in the subcomponent appears higher than in the rest of the cluster.

\subsection{Vicinity of A2111}

	The combination of the PSPC and RHRI observations allows us
to detect extended sources as candidates for additional clusters in the 
vicinity of A2111. It turns out that the brightest PSPC source
($15^h39^m1\fs2, +34\arcdeg36\arcmin 8\arcsec$), 
besides A2111, in our examined region (Fig. 1) is such a candidate.
The source is extended at the confidence level of 99.7\%, has a
count rate of $7.6\pm1.2 \times 10^{-3} {\rm~counts~s^{-1}}$, and
has a hard spectrum consistent with a thermal 
plasma of several keV.  The source can be barely seen as an extended feature 
in the RHRI image (Fig. 2). This weak contrast of the source 
is due to both the high non-cosmic X-ray background and the degraded RHRI
PSF near the edge of the image. The integrated flux within a 
30$^{\prime\prime}$ radius is consistent with that inferred from
the PSPC rate. 

	Furthermore, the X-ray source is apparently associated with a compact 
group of optical objects
(Fig. 2), which are classified as either galaxies or multiple objects in
the APM catalog. The source is thus likely a cluster (the NW cluster
hereafter). Three major optical objects in the group appear a bit
brighter than galaxies in A2111, indicating that the group is probably not
as distant as A2111. However, if this cluster is at the 
same redshift as A2111, the implied 0.1-2~keV luminosity is then 
$\sim 3 \times 10^{43} h^{-2}_{50}{\rm~ergs~s^{-1}}$ (or about
4\% of the A2111's), reasonable for the apparent optical
richness. Incidentally, the cluster is 
projected in a position to which A2111 directs its subcomponent's tail.

\section{Discussion}

\subsection{General Consideration}

	Can A2111 be a chance superposition of two noninteracting clusters?
To produce the observed off-center morphology, A2111 needs a superimposed
cluster that is 2-3 times brighter than that of the NW cluster. The statistical
probability for such a superposition is small ($\sim 10^{-2}$). Furthermore, 
if the superimposed cluster has an approximately circular or elliptical 
morphology, we would still have difficulties in reproducing the tail of the 
subcomponent. In addition, based on the redshift measurements of 24
objects in the A2111 field, Lavery \& Henry (1986) find no strong
evidence for a superposition of a foreground or background cluster
with A2111.  Six field galaxies were identified at five distinct
redshifts. Therefore, the subcomponent is most likely intrinsic to A2111.
	
\subsection{A2111 as a Head-on Subcluster Collision}

	Our results have strong implications on the dynamic state of A2111. 
First, as argued by
Mohr, Fabricant, \& Geller (1993), the shift of X-ray image centroid
indicates that the center of mass of the cluster is changing as a function 
of scale and that the cluster is not a relaxed, equilibrium system
(see also Ulmer, McMillan, \& Kowalski 1989).  
Second, our identification of the comet-shaped subcomponent suggests
that this unstable state is most likely the result of a major subcluster
merger. Third, the approximate alignment among the subcomponent 
elongation, the galaxy distribution,
and the relative orientation of the two major galaxies 
indicates that the merger is a nearly head-on collision.
The closeness of the two major 
galaxies, which may well be the central galaxies of the two merging 
components, is consistent with a recent core impact.

	By comparing our observed characteristics of A2111 
with N-body/hydrodynamic simulations of 
such collisions (Schindler \& M\"uller 1993; Roettiger, Burns, \& Loken 1993; 
Pearce, Thomas, \& Couchman 1994), we can learn more about the 
merger history and process. According to the simulations, 
the ICM can have been heated by the passage of the subcluster, 
consistent with the flat X-ray distribution
of the main component in the northwestern part of A2111 (\S 3.3). 
A shock may have developed during the core impact. The
present position of the shock front is presumably represented by 
the compression at the southeastern edge of the subcomponent blob and 
by the fan-shaped contours on both sides of the blob, indicating
an invasion of the subcluster from the northwest along the major axis.

	This merger scenario also naturally explains 
the morphology and thermal state of the subcluster, as characterized by 
the subcomponent. The subcluster, forcing into the main cluster, 
should have left a trail of X-ray-emitting gas relics, which might 
be responsible for the subcomponent tail. The simulations show that the 
ICM inside the subcluster, emerging out from the core of the main cluster,
should be heated to a temperature of $\sim 10^8$~K.   
Consistent with being so hot (\S 3.4), the blob of 
the subcomponent may represent the core of the subcluster. The 
ram-pressure confinement of the snow-plowing subcluster accounts for 
the high X-ray intensity of the blob. 

	The lobe in front of the blob is very intriguing, however. If the
subcluster is now falling back to the main cluster center, the residual
of the subcluster may then explain the lobe. But, totally absent
in the PSPC soft band (Fig. 5), the lobe probably has a hard spectrum,  
inconsistent with the prediction of a low temperature $\sim 5 \times 10^6$~K
(e.g., Roettiger, Burns, \& Loken 1993). Alternatively, the feature may 
represent a separate subcluster. Fig. 6 indeed shows a condensation
of galaxies in the region. But the true nature of the feature remains 
uncertain. 

	 The average ICM temperature of A2111, 3.1(2.1-5.3)~keV (\S 3.4), 
is low, compared to clusters of comparable optical richness  (e.g., 
Edge \& Stewart 1991). The Coma cluster  ($N_{30\%}=94$; $f_b = 0.03\pm0.01$)
is an example. The temperature $9.1 \pm 0.7$~keV of this nearby cluster
(Hughes, Gorenstein, \& Fabricant 1988) is about a factor of 2 higher 
than that of A2111. As a distant example, the well-known cluster 
CL0016+16  ($N_{30\%}=65$; $f_b = 0.02\pm0.07$) at $z=0.54$, 
has a temperature of $8.2 (6.7-10)$~keV (Neumann \& B\"ohringer 1996
and references therein). 

	The relatively low ICM temperature of A2111 may be directly related to 
the ongoing major merger. Individual components before the merger 
can be considerably cooler. Relatively cool gas originally on
the outskirts of the components may have penetrated into 
the central region of the current system, resulting in enhanced
soft X-ray radiation. Furthermore, if A2111 is of multi-temperatures and
if a pressure balance approximately holds, a cooler component tends
to have a higher emission measure. The soft emission from this
cooler component, dominating in the \rosat band, could have resulted in 
an underestimate of the average ICM temperature. An upcoming {\sl ASCA}
observation will, we hope, provide more conclusive 
data on the multi-temperature state of the cluster. 

	Our data indicate an absence of a cooling flow in A2111. The
head-on collision could have destroyed such a flow, if existing previously
(Roettiger, Burns, \& Loken 1993). Currently, the cooling
time scale at the X-ray emission peak, for example, is $\sim (2
\times 10^{10} {\rm~yr}) h_{50}^{-1/2} (T_e/10^{8}~{\rm K})^{1/2}$, likely
longer than the age of the cluster.

	Strong evidence for major subcluster mergers has been observed
in two nearby clusters: A2256 ($z = 0.060$; Briel et al. 1991; 
Briel \& Henry 1994; Markevitch 1996)
and A754 ($z = 0.054$; Henry \& Briel 1995; Henriksen \&  Markevitch 1996). 
The overall X-ray morphology of A754 resembles that of A2111. But, 
the off-center emission peak in A754 appears as a bar nearly perpendicular 
to the overall elongation of the cluster, and has a temperature ($\sim 3$~keV) 
lower than the cluster average ($\sim 9$~keV). 
Henriksen \&  Markevitch (1996) show that the X-ray surface 
brightness and temperature distributions of A754 are consistent with
a non-head-on merger (e.g., Evrard,  Metzler, \& Navarro, 1996).
This scenario may also explain the large separation, 12$^\prime$ 
(a projected distance of $\sim 1$~Mpc; Henry \& Briel 1995),
between the two major galaxies in the cluster.  
A similar interpretation may also hold for A2256.
The subcomponent of A2256 is cooler ($\sim 4$~keV) 
than the surrounding medium ($\sim 8$~keV),
and has a distinct compression of contour lines in the direction 
approximately perpendicular to the major axis  of the 
cluster. This compression is presumably due 
to the snowplow effect on the leading edge of the subcluster (Briel et al. 
1991), indicating that the subcluster is {\sl spiraling} into A2256.
Therefore, the impact appears to be a non-head-on collision as well. 
In comparison, the orientation of the subcomponent 
in A2111 is nearly parallel, instead of perpendicular, to the major axis 
of the cluster in both X-ray and optical, and the gas temperature of the 
subcomponent appears hotter, instead of cooler, than the cluster average. 

\subsection{The Subcluster Merger and the Blue Galaxy Population}

A2111 contains a high fraction ($f_b = 0.16\pm0.03$) of blue galaxies, 
as measured both photometrically (BO) and spectroscopically (Lavery \& Henry 
1986). Lavery \& Henry (1988, 1994) further show that the large
majority of the blue galaxies in A2111 have disk-like morphology and that   
star formation is extended over galactic disks in two nearly face-on spiral
galaxies. Could the ongoing merger be responsible?

	 An ongoing merger is expected to contain a larger fraction of
spiral galaxies  than a relaxed cluster of the same mass. 
Generally, the spiral fraction $f_{sp}$ of a cluster is known to 
decrease with increasing X-ray luminosity (or the richness; Bahcall, 1977); 
for nearby clusters, most of which are more-or-less relaxed systems,  $f_{sp} 
\approx 10^{15} L_{bol}^{-0.35}$ and
$L_{bol} \propto  \sigma^{2.90}$, where $\sigma$ is the galaxy 
velocity dispersion (Edge \& Stewart 1991). Because the number of
galaxies in a relaxed cluster $N_g \propto \sigma^2$, we have $f_{sp} \propto
N_g^{-0.51}$. Thus, an average nearby cluster with an X-ray luminosity 
of A2111's (\S 3.4) is expected to have $f_{sp} \sim 0.16$.  For a cluster
at the redshift of A2111,  we assume a more general form $f_{sp} 
\propto N_g^{-\gamma}$, where $f_{sp}$ may also include late-type galaxies
besides spirals. Simple algebra shows that a merger of two relaxed subclusters 
results in a system of an initial spiral fraction 
$f^m_{sp} \approx f_{sp} (1+r_g^{1-\gamma})/(1+r_g)^{1-\gamma}$, where
$r_g$ is the galaxy number ratio of the subclusters.
When the two subclusters have the equal number of galaxies (i.e., $r_g = 1$),
$f^m_{sp} = 2^\gamma f_{sp}$, which is the maximum if $\gamma < 1$ (or
the minimum if $\gamma > 1$). If $\gamma \sim 0.51$, for example, the 
$f^m_{sp}$ value of A2111 is then about 0.23. This spiral fraction can
be substantially larger, if $\gamma$ is greater and/or if the merger 
involves more than two subclusters. However,
no all spirals are blue galaxies as defined by BO.

	The merger may further trigger starbursts in some of the
spirals, accounting for the large fraction
of blue galaxies observed in A2111.  The starbursts could result
from an enhancement of two-body interactions in the infalling
subcluster (Lavery \& Henry 1988: Lavery, Pierce, \& McClure 1992),
from frequent encounters with cluster members, 
together with the tidal forces of the mean cluster potential (Moore et al. 
1996; Henriksen \& Byrd 1996), and/or from the high pressure of the ICM. 
The blue galaxies of A2111 show a higher velocity 
dispersion, by about 50\%, than the red cluster members (Lavery 1988), 
consistent with the ongoing merger scenario. The cluster environmental
effects can also gradually transform late-type galaxies to early-type ones 
(Oemler et al. 1997), resulting in the decrease of the
spiral fraction with time after a merger.
	
	In an ongoing study of other distant ($z \gtrsim 0.2$) 
BO clusters observed with the
\rosat PSPC (Wang \& Ulmer 1997 in preparation), we find that 
clusters of high $f_b$  consistently show 
relatively low average ICM temperatures, contain substantial amounts of
substructure, and tend to be strongly elongated. Thus a high $f_b$ value of 
a cluster is an indicator of a dynamically young system. Furthermore,
according to the hierarchical clustering
theories of the structure formation, an observed high-$z$ rich cluster
is typically a younger system and is assembled
from units of smaller masses than a present-day cluster of the same mass. 
Therefore, the increasing 
$f_b$ of rich clusters with $z$ --- the Butcher \& Oemler effect --- is a
natural consequence of the hierarchical structure formation process 
(Kauffmann 1995).

\section {Summary}

Based on our \rosat PSPC and RHRI observations, plus a deep R-band CCD image,
we have conducted a detailed spatial and spectral analysis of the richest 
BO cluster A2111. Our analysis includes a 2-D morphological characterization 
of the cluster, using a moment method improved to remove background effects. 
We find that the cluster has a strongly elongated morphology,
similar in optical and X-ray, and that both the centroid and ellipticity 
of the X-ray morphology vary significantly with scale.

	We conclude that A2111 contains at least two major X-ray-emitting 
components. The main component, characterized by the northwestern half of 
the cluster's X-ray emission, has a flat distribution,
whereas the subcomponent is more-or-less comet-shaped. The
orientation of the subcomponent aligns well with the elongation of
both the galaxy and the X-ray surface brightness distributions
of the cluster.
The gas in the subcomponent appears hotter than in the main component. 
The average ICM temperature of A2111 is, however, unusually low (by
a factor of $\sim 2$), compared to clusters of similar 
optical richness.

	The characteristics of the subcomponent suggest that
A2111 represents a head-on merger between two clusters. Numerical 
simulations of such mergers show similar morphological and spectral 
features as observed in A2111, if the core impact has just occurred. 
This ongoing merger is also likely responsible for the 
high blue galaxy fraction of the cluster. 

\acknowledgements
	
	We thank Mike Harvanek for his helping with classifying CCD objects,
J. Stocke for his involvement in the proposing of the PSPC observation, and
the referee for useful comments. This project is supported by NASA under 
Grant 5-2717.
\begin{table}[htb]
\begin{tabular}{lcccc} 
\multicolumn{3}{c}{\bf Table 1}\\
\multicolumn{3}{c}{\bf $\beta$ Model Characterization of the
X-ray Surface Brightness Profiles\tablenotemark{a}}\\ [0.1in]
\hline\hline 
Parameter 	& PSPC & RHRI\\
\hline      
$\beta$ 	& 0.54 (0.50-0.59) & 0.47 (0.44-0.52)\\
$r_c$ ($h^{-1}_{50}$~Mpc)& 0.21(0.17 - 0.26)&0.18(0.15 - 0.22) \\
$I_{o} (10^{-2} {\rm~cts~s^{-1}~arcmin^{-2}}$) &3.2(2.7-3.8)&1.5(1.3-1.7)\\
$n_{e,o}$($10^{-3} h_{50}^{1/2}{\rm~cm^{-3}}$) &3.9 (3.4-4.6)&4.1 (3.6-4.7)\\
$\chi^2/d.o.f$ 	& 47.8/36	& 72.5/68\\
\hline 
\end{tabular}
\tablenotetext{a}{Uncertainties in parameter values,
as presented in parentheses, are all at the 90\% confidence level. 
The derivation 
of the central electron number density $n_{e,o}$ assumes an oblate shape
of the X-ray-emitting medium and the best-fit Raymond \& Smith model of
the integrated PSPC spectrum of A2111 (\S 3.4).
}
\end{table}
\vfil


\vfil
\clearpage

\begin{figure} \caption{PSPC image of A2111 in the 0.5-2~keV band. 
The contours represent the distribution of the X-ray surface brightness, 
which has been  
corrected for exposure and has been smoothed adaptively with a Gaussian,
the size of which is adjusted at each pixel to achieve a count-to-noise 
ratio of 4. Each contour is 2$\sigma$ 
(50\%) above its lower level; the lowest contour
is at $3.4 \times 10^{-4} {\rm~counts~s^{-1}~arcmin^{-2}}$. 
\label{fig1}}
\end{figure}

\begin{figure} \caption{RHRI surface brightness contours overlaid on the 
digitized optical sky survey image of the A2111 field. The X-ray 
surface brightness distribution has been smoothed with a Gaussian of size
30$^{\prime\prime}$ (FWHM), comparable to the RHRI PSF at corners
of the field. The contours are at 1.4, 2.0, 2.6, 3.2, 
3.8, 5.0, 6.2, 7.4, 8.6, 9.8, 11, and 12 $\times 10^{-3} 
{\rm~counts~s^{-1}~arcmin^{-2}}$.
{\sl Crosses} mark point-like X-ray sources detected outside the central 
3\farcm5 radius.
\label{fig2}}
\end{figure}

\begin{figure} \caption{PSPC image of A2111 in the 0.5-2~keV band after
point-like X-ray sources are excised. Individual dots represent PSPC counts in 
the 0.5-2~keV band. The regions within 90\% source 
flux-encircled radii are replaced by randomly generated counts with intensity 
interpolated from neighboring average. The rest is the same as in Fig. 1.
\label{fig3}}
\end{figure}

\begin{figure} \caption{RHRI intensity contours overlaid on the R-band  
image of A2111. The X-ray surface brightness 
is adaptively smoothed to achieve a count-to-noise ratio 
of $\sim$ 6 over the image. The contours are at 1.2, 2.1, 3.2, 4.8, 6.9, 9.7,
13, 18, and $25 \times 10^{-3} {\rm~counts~s^{-1}~arcmin^{-2}}$. 
\label{fig4}}
\end{figure}

\begin{figure} \caption{PSPC surface brightness contours 
in the 0.14-0.28~keV band (upper panel)
and in the 0.5-2~keV band (lower panel) superposed on the R-band CCD 
image of A2111. The soft band contours are at 0.75, 1.1, 1.7, 2.5, 
and 3.8 $10^{-3} {\rm~counts~s^{-1}~arcmin^{-2}}$. The rest is the same 
as in Fig. 1.
\label{fig5}}
\end{figure}

\begin{figure} \caption{Distribution of galaxies in the central field  of 
A2111. The galaxies, represented by {\sl pluses}, are selected 
in the magnitude 
range of 17 to 23.7. The contours illustrate the distribution after being
adaptively smoothed with a Gaussian filter of a count-to-noise ratio 
$\sim 4$, and are at 12, 18, and $27 
{\rm~galaxies~arcmin^{-2}}$. Two main galaxies in the central region
are marked as {\sl diamonds}.
\label{fig6}}
\end{figure}

\begin{figure} \caption{Radial surface brightness profiles of A2111 derived
from the PSPC observation (a) and the RHRI observation (b). 
The data are represented by {\sl crosses} with  standard 
error bars. The curves represent the best fits of the
$\beta$ model to the profiles. 
\label{fig7}}
\end{figure}

\begin{figure} \caption{Characterization of the 2-D X-ray morphology of A2111
with ellipses on scales from 1\arcmin-8\arcmin. The triangle marks the 
global X-ray centroid determined on the largest scale. 
The dots represent PSPC counts as in Fig. 3.
\label{fig8}}
\end{figure}

\begin{figure} \caption{Center coordinates, ellipticity, and orientation 
of the ellipse as a function of semi-major axis. The coordinates
are relative to the central galaxy position 
$15^h39^m40\fs4, +34^d 25^{\prime} 27\arcsec$.
Error bars are at the 
90\% confidence level. Note that individual measurements 
are not statistically independent (see text).
\label{fig9}}
\end{figure}

\begin{figure} \caption{PSPC surface brightness profile as a function of 
the semi-major axis in the main-component elliptical 
coordinates (see text). The rest is the same as in Fig. 7.
\label{fig10}}
\end{figure}

\begin{figure} \caption{PSPC surface brightness residuals, 
after subtracting the elliptical $\beta$ model from Fig. 6. The contours 
are at -1.2, -0.6, -0.3, 0.3, 0.6, 1.2, 
2.4, 4.8, 9.6, and 19.2 $\times 10^{-3} {\rm~counts~s^{-1}~arcmin^{-2}}$. 
The two main galaxies are marked as diamonds. The ellipse illustrates the
ellipticity and orientation of the main component.
\label{fig11}}
\end{figure}

\begin{figure} \caption{PSPC spectrum of A2111 and
the best-fit Raymond \& Smith model (histogram).
\label{fig12}}
\end{figure} 


\clearpage

\begin{references} 
\reference{} Briel, U. G., \& Henry, J. P. 1994, Nature, 372, 439
\reference{} Briel, U. G., et al. 1991, A\&A, 246, L10
\reference{} Buote, D. A., \& Canizares, C. R. 1994, ApJ, 427, 86
\reference{} Butcher, H., Oemler, A., \& Wells, D. C. 1983, ApJS, 52, 183
\reference{} Butcher, H., \& Oemler, A. 1984, ApJ, 285, 426 (BO)
\reference{} Carter, D., \& Metcalfe, N. 1980, MNRAS, 191, 325
\reference{} Cavaliere, A., \& Fusco-Femiano, R. 1976, A\&A, 49, 137
\reference{} Couch, W. J., Ellis, R. S., Sharples, R. M., \& Smail, I. 1994, ApJ, 430, 121
\reference{} Dressler, A., \& Shectman, S. A. 1988, AJ,  95, p85
\reference{} Dressler, A., \& Gunn, J. E. 1992, ApJS, 78, 1
\reference{} Edge, A. C., \& Stewart, G. C. 1991, MNRAS, 252, 414
\reference{} Efron, B., \& Tibshirani, B. J. 1993, in An Introduction to
the Bootstrap, Champman \& Hall, New York
\reference{} Evrard, A. E., Metzler, C. A., \& Navarro, J. F. 1996, ApJ, 469, 494
\reference{} Fabricant, D. F., Bautz, M. W., \& McClintock, J. E. 1994, AJ, 107, 8
\reference{} Geller, M. J., \& Beers, T. C. 1982, PASP, 94, 421
\reference{} Hasinger, G., et al. 1994, Legacy, 4, 40
\reference{} Henry, J. P., \& Briel, U. G. 1995, ApJL, 443, 9
\reference{} Henriksen, M. J., \& Byrd, G. 1996, ApJ, 459, 82
\reference{} Henriksen, M. J., \& Markevitch, M. L. 1996, ApJL, 466, 79
\reference{} Hughes, J. P., Gorenstein, P. D., \& Fabricant, D. 1988, ApJ, 329, 82
\reference{} Jones, C., \& Forman, W. 1992, in Clusters and Superclusters
of Galaxies, ed. A. C. Fabian (Dordrecht: Kluwer), 49
\reference{} Kauffmann, G. 1995, MNRAS,  274, 161
\reference{} Kuerster, M. 1993, {\sl ROSAT} Status Report, No. 67
\reference{} Lavery, R. J. 1988, PhD. thesis, University of Hawaii
\reference{} Lavery, R. J., \& Henry, J. P. 1986, ApJL, 304, 5
\reference{} Lavery, R. J., \& Henry, J. P. 1988, ApJ, 330, 608
\reference{} Lavery, R. J., \& Henry, J. P. 1994, ApJ, 426, 524
\reference{} Lavery, R. J., Pierce, M. J., \& McClure, R. D. 1992, AJ, 104, 2067
\reference{} Markevitch, M. 1996, ApJL, 465, 1
\reference{} McMillan, S. L. W., Kowalski, M. P., \& Ulmer, M. P. 1989, ApJS, 70, 723
\reference{} Mohr, J. J., Fabricant, D. G., \& Geller, M. J. 1993, ApJ, 413, 492
\reference{} Mushotzky, R. F., et al. 1996, ApJ, 466, 686
\reference{} Neumann, D. M., \& B\"ohringer, H. 1996, preprint astro-ph/9607063
\reference{} Newberry, M. V., Kirshner, R. P., \& Boroson, T. A. 1988, ApJ, 335, 629
\reference{} Nichol, R., Ulmer, M. P., Kron, R. G., Wirth, G. D. \& Koo, D. C.
1994, ApJ, 432, 464
\reference{} Bahcall, N. A. 1977, ApJL, 218, 93
\reference{} Oemler, A., Dressler, A., \& Butcher, H. R. 1997, ApJ, 474, 561
\reference{} Pearce, F. R., Thomas, P. A., \& Couchman, H. M. P. 1994, MNRAS, 268, 953
\reference{} Roettiger, K., Burns, J., \& Loken, C. 1993, ApJ, 403, L53
\reference{} Sandage, A., Kristian, J., \& Westphal, J. A. 1976, ApJ, 205, 688
\reference{} Sarazin, C. L. 1980, ApJ, 236, 75
\reference{} Schindler, S.,  \& M\"uller, E. 1993, A\&A, 272, 137
\reference{} Snowden, S. L., McCammon, D., Burrows, D. N., \& Mendenhall, J. A.  1994, ApJ, 424, 714
\reference{} Stark, A. A., et al. 1992, ApJS, 79, 77
\reference{} Tomita, A. et al. 1996, AJ, 111, 42
\reference{} Tr\"umper, J. 1992, QJRAS, 33, 165
\reference{} Ulmer, M. P., McMillan, S. L. W., \& Kowalski, M. P. 1989, ApJ, 338, 711
\reference{} Yee, H. K. C. 1991, PASP, 103, 396
\reference{} Zabludoff, A. I., et al. 1996, ApJ, 466, 104
\end{references}
\end{document}